\documentclass[10pt]{article}
\usepackage{epsfig}
\usepackage{amsmath}
\usepackage{amsfonts}
\usepackage{hyperref}
\usepackage{float}
\usepackage{lscape}

\usepackage[T1]{fontenc}
\usepackage[utf8]{inputenc}

\begin{document}
\newcommand{\NP}{N\!P}
\newcommand{\Vh}{V_h}
\newcommand{\Eh}{E_h}
\newcommand{\Vl}{V_l}
\newcommand{\El}{E_l}
\newcommand{\Vs}{V_s}
\newcommand{\Vt}{V_t}
\newcommand{\Vi}{V_i}
\newcommand{\card}[1]{|#1|}
\newcommand{\outd}[1]{d^+(#1)}
\newcommand{\ind}[1]{d^-(#1)}
\renewcommand{\th}{\mbox{$^{\scriptstyle {\rm th}}$}}
\newtheorem{mydef}{Definition}
\pagestyle{empty}
\vspace*{10mm}

\begin{center}
 {\Large \bf An Experimental Evaluation of List Coloring Algorithms}

\vspace{7mm}

 {\large \bf Andrew Ju and Patrick Healy\\}

\vspace{1mm}
 {\large Department of Computer Science,\\
  University of Limerick, \\Limerick, IRELAND\\
  \texttt{a.ju@acm.org}\\}
\end{center}

\vspace{.1cm}
\noindent
  \begin{center}
  \textbf{Abstract}
  \end{center}
  
  The \emph{list coloring} problem is a variant of vertex coloring where
  a vertex may be colored only a color from a prescribed set.  Several
  applications of vertex coloring are more appropriately modelled as
  instances of list coloring and thus we argue that it is an important
  problem to consider.  In spite of its importance few published
  algorithms exist for list coloring, however.  In this paper we review
  the only two existing ones we could find and propose a \textit{branch-and-bound} 
  based one.  We conduct an experimental evaluation of those algorithms.

\section{Introduction}\label{sec:intro}%

In the classical vertex coloring problem one asks if one may color the
vertices of a graph $G = (V, E)$ with one of $k$ colors so that no two adjacent
vertices are similarly colored; the corresponding optimization problem
seeks to find the minimum value $k$ for a graph that admits a legal $k$
coloring.  In 1976 Vizing~\cite{vizing-76:list-col} proposed an
additional restriction on the coloring by supplying for each vertex a
list of permissible colors (in this paper we refer it as the color availability list).

Many problems that rely on vertex coloring might be modelled more
appropriately using list coloring.  For example, exam timetabling is
frequently modelled as a vertex coloring problem where graph edges
represent subjects that may not be scheduled simultaneously.  Other
constraints, such as preference for the times an exam may be scheduled,
are often considered to be \emph{soft constraints} \cite{QuBur2009}.  By
supplying a list of (in)appropriate hours for each exam one may model
the problem more accurately as an instance of list coloring.  Similarly,
the frequency assignment problem for cellular telephone networks and
\textsc{WLAN}s may be modelled more accurately by restricting the
coloring of vertices (transmitters or routers) to a specified set
\cite{borndorfer-etal-98:FreqAss}.

The list coloring problem is defined to be the search for a proper
vertex coloring of a graph of minimum size such that each vertex is
colored a permissible color.  Thus, an instance of the \emph{list
  coloring problem} is a graph accompanied by a list
  attached to each vertex of at most length
  $n$.

Closely related to the list coloring problem is the \emph{weighted vertex
  coloring} or \emph{vertex multicoloring} problem
\cite{malaguti2010,mehrotra-trick-07:multicol-BP,journals/dmgt/CaramiaF04,NET:NET1028}.
In this problem each vertex has a weight associated with it and the
graph $G = (V, E)$ must be multicolored so that vertex $v \in V$ is assigned a
\emph{set} of colors $C_v$ where $C_u \cap C_v = \emptyset,  \forall (u,v) \in E$;
the objective is to find $\chi(G,w)$, the minimum number of colors required
to color $G$ and that satisfies the vertex weight requirement.  When,
further, each vertex is supplied with a list of permissible colors one
arrives at the \emph{list multicoloring} problem
\cite{DBLP:journals/corr/abs-1202-4842}.

Clearly the list coloring problem is as hard as vertex coloring, for the
latter reduces to the former (in polynomial time) through supplying, for
every vertex, all colors as its permissible list.  Few published
algorithms exist for the list coloring problem in its most general form.
In the context of frequency assignment Borndörfer \emph{et al.} \cite{borndorfer-etal-98:FreqAss}  present
several heuristics that incorporate problem-specific requirements.
Likewise with Garg \emph{et al.}  \cite{Garg96distributedlist} though
their interest is in developing a distributed solution.

In this paper we investigate the performance of three list coloring
algorithms.  The first we consider is the greedy, random algorithm
$k$-GL (Greedy List) proposed by Achlioptas and Molloy
\cite{Achlioptas97theanalysis}.  Our second algorithm from the
literature is a maximal independent set-based heuristic algorithm
\cite{tsouros-satratzemi-05}.  Finally we propose a new
branch-and-bound based algorithm \textsc{ELC}.  While the running time of the
latter cannot be expected to be competitive with the former two it does
provide a useful reference point against which one may consider their
performance.

In the following section we describe the three algorithms we have
implemented and the context of our experiments.  Following that, in
Section~\ref{sec:expers} we provide the outcome of our experiments.  In
Section \ref{sec:concls} we conclude the paper and suggest areas of
further research.

\section{Experimental Context}\label{sec:exper-context}%

We implemented two algorithms from the literature as well as a newly
developed branch-and-bound algorithm \texttt{ELC}.  Of the two previously published
algorithms one is randomised and the second is deterministic; both are
heuristic.  We describe these algorithms in the following sections.

\subsection{The $k$-GL algorithm}\label{sec:k-gl-alg}

Achlioptas and Molloy propose a greedy algorithm they call
$k$-Greedy-List or $k$-GL \cite{Achlioptas97theanalysis}.  Each vertex
is supplied with a permissible color list $L_v = \{1, 2, 3, ..., k\}$
(the contiguous sequence of integers between 1 and $k$) in order to
facilitate their analysis.  However, it would not be difficult to modify
the algorithm in order to cater for a) lists of varying lengths and, b)
non-contiguous sequences.

The algorithm proceeds by picking a vertex $v$ that is deemed most
critical as measured by the number of remaining colors on its
permissible list, $L_v$, with ties broken randomly.  If set $L_v$ is not
empty, a color randomly chosen from it is assigned to $v$ and since that
color can no longer be used in $v$'s neighbourhood it is removed from
each neighbour's permissible list; if $L_v$ is empty, then the
algorithm fails in finding a solution.

The algorithm we implemented is a modification
of the original \cite{Achlioptas97theanalysis} so that it accepts lists
of non-contiguous sequences.  We discuss its performance in Section
\ref{sec:heur-evaluation}.

\subsection{The $LC$ algorithm}

Tsouros and Satratzemi \cite{tsouros-satratzemi-05} propose a deterministic 
heuristic algorithm that centres around finding a maximal
independent set $S$ where all vertices in $S$ share a common permissible color 
$c$ at each iteration.  Since the subgraph induced by the independent set $S$ is 
edgeless all vertices in the set $S$ may be colored with the common color $c$.

The algorithm proceeds by determining $Q_i = \{q_{i_1}, q_{i_2}, ...\}$ the set of vertices that can
be colored $i$ ($v \in Q_i \Leftrightarrow i \in L_v$) and ordering elements of $Q_i$
in a way that $deg(q_{i_j}) \leq deg(q_{i_{j+1}})$ (line 5-6).  Then amongst
all those $Q_i$s a search is made for $S$, the largest maximal
independent set (line 10; code not shown); the vertices in $S \subseteq
Q_j$ are colored $j$ and color $j$ is removed from $LL$ ($LL =
\cup{}_{i=1}^{|V|}L_i$).  Data structures are then updated appropriately.
The algorithm fails in finding a solution if $LL$ is empty when there
are still uncoloured vertices. The algorithm makes little effort to
compute a ``good'' maximal independent set: it initialises the independent set $S$ by 
adding the first element in $Q_i$, and then performs a linear scan over the remaining 
elements in $Q_i$,  if the element is not in $S$'s neighbourhood, it is then being added in $S$.  
There are three nested loops (one not shown) with each iterating over, at worst, the 
set of vertices and set of colors, and thus the overall running time is $O(n^3)$.
We report on its performance in Section \ref{sec:heur-evaluation}.

\subsection{ELC -- New Algorithm}\label{sec:aelc}

We developed a branch-and-bound (BB) based algorithm, ELC.  With sufficient time ELC will find the minimum
coloring subject to the supplied list constraints for each vertex.  At each step the algorithm
selects the ``next up'' (most critical) vertex and considers all of its
permitted colors in turn. If all vertices are colored the
search has reached a leaf node of the search tree, and therefore a feasible coloring has been found. If the total
of colors used is equal to the lower bound, the search is terminated as the optimal solution is found;
otherwise, if the number of colors used to date exceeds that of a
previous feasible solution (named UB -- upper bound) then the search path is
abandoned and another color possibility is examined.  If the number is
less, then a new upper bound is determined.  This upper bound is used to
prune the search tree if the total colors used in the current coloring
process is greater than or equal to the current upper bound.  This makes
ELC a branch and bound algorithm.

In the following section we discuss the heuristics that we have
investigated for determining the next up vertex for coloring and also describe
the initial coloring procedure that we have determined to be useful in a
branch-and-bound setting. We evaluated its performance in Section
\ref{sec:elc}.

\subsubsection{Branch-and-bound Issues}

At each step, ELC picks the ``next up'' vertex from the uncolored
subgraph and colors it accordingly.  It is clear that with a \textit{proper}
coloring order (the sequence of picking each vertex for coloring and assigning color to it), the algorithm will arrive at a tighter upper bound earlier.
Therefore, a proper vertex selecting rule shall be deployed in ELC as it can pick the
next up vertex for coloring more carefully.

\begin{mydef}\label{def:sd}
Given a simple undirected graph $G = (V, E)$, color availability list $L_v = \{c_1, c_2, ..., c_k\}$ for each vertex $v \in V$ and $C$ a partial coloring of $G$ vertices. We define the \textbf{unavailable degree} of a vertex as the number of different colors which are unavailable (either not in its color availability list or have been used on any adjacent colored vertex).
\end{mydef}

We considered a very effective yet simple vertex selecting rule for vertex
coloring, DUA-h\footnote{Note that this is similar with the one in \cite{Bre79} but not 
the same: in \cite{Bre79} the saturation degree is ``the number of different 
colors to which it is adjacent (colored vertices)'', here this term is redefined 
as in Definition \ref{def:sd}. The performance is largely improved with this
new definition. Due to space, comparison of result is eliminated for this paper.}.  We
denote $deg_{ua}{v}$ the \emph{unavailable} degree as defined in Definition 
\ref{def:sd} and $deg_u(v)$ as the number of vertices in the uncoloured 
subgraph of $G$ to which vertex $v$ is adjacent. The vertex selection rule is then described as:  at each step pick the uncolored vertex $v \in V$ where $deg_{ua}(v)$ is maximal; 
if there a tie then pick the vertex amongst the set of ties where $deg_u(v)$ is maximal;
if there is still a tie after the second comparison, 
pick a vertex $v$ amongst these ties lexicographically.
The ELC algorithm we present in this section is based on this vertex selection rule.

Prior to implementing the
ELC algorithm, various heuristics for finding proper initial upper
bounds were investigated. During our investigation, we found that by
simply providing a tighter upper bound initially will not make difference 
in later stage since the algorithm simply can not reach the corresponding
leaf node that confirms the bound. With regard to the initial lower bound, we simply
use size of the clique as the bound; this is also mentioned below. 

When branching the search tree, the algorithm traverses on the very left branch of 
the tree and tries to reach the leaf node with a result of ``either a feasible solution or
no solution'', and then backtrack in the search tree to reach its parent node,
and start branching the next branch with another permissible color. Therefore, the very left
branch is critical to the whole search as it affects almost all the remaining searches, we call this the ``search order'' of the tree. Now the question is, how to obtain a good search order. For this, ``initial coloring'' is the answer we suggest. This is discussed further below.   

\subsubsection{Initial coloring}

In vertex coloring problems, both DSATUR\cite{Bre79} and PASS algorithms \cite{PASS} start
with the procedure of trying to find a clique as large as possible and then color the clique as the initial
coloring. The initial coloring of the clique can generate initial data which can be used to determine the next 
up vertex for coloring and therefore make the search order better. For the same purpose, we use clique coloring
as the initial coloring in list coloring, and also the clique size can be used as the lower bound for the BB search \footnote{This is a very weak bound, but to best of our knowledge there is no alternatives at this moment}. In vertex coloring since 
each color is identical if the coloring is given in a tight manner, the assignment of each color 
to each of the vertices in the clique can be done lexicographically, but in list coloring
each of the colors is different from each other (due to the color availability list), the assignment can
not be done lexicographically, and to keep the solution completeness, we need to make sure 
the use of the initial clique coloring doesn't miss any potentially optimal results.  

In list coloring, each of the feasible clique colorings can be used as an initial coloring, and 
each of the initial colorings can be regarded as the ``early level'' branches of the BB tree after pruning the
unnecessary branches. Now, since each of the initial colorings are different, and yet we don't know which 
clique coloring may lead to the branches where optimal colourings exist,   a safe way is all of them should be used as the initial coloring. But the reality is, depending on the graph size, it could still be difficult to complete the BB search with the initial coloring; this can be evidenced by our initial investigation which shows that ELC only managed to complete a small portion of feasible clique colorings.  To avoid the case mentioned above, therefore in this
paper we implemented the algorithm as for each the upcoming BB search there is a limitation of 5,000 iterations. With this, we managed to let the algorithm try each of the possible clique coloring to benefit from the initial clique coloring, therefore we didn't miss any ``good'' branches. 

\paragraph{Double clique coloring} Seeing the improvements made by doing an initial single clique coloring, we extend the idea by applying an extra clique coloring(therefore, find two cliques first, and then use the clique coloring as the initial coloring to conduct the BB search), this will 1) prune more unnecessary branches before the BB search, 2) give more information for selecting the next up vertex to avoid ties, and 3) provide a tighter lower bound\footnote{The minimal number of colors required to color the two cliques is then used as the LB}. Based on this idea, we implemented \texttt{dcc} - another variant of ELC. The result from our investigation shows that this is a useful technique that can reduce the gap between the heuristic result and the optimal. The result is reported and discussed in Section \ref{sec:expers}.

\section{Experimental Evaluation}\label{sec:expers}%

To facilitate a preliminary experimental analysis of the algorithms it
is reasonable that we focus on small graph sizes $|V|$ and list lengths $k$.

For tested graphs we used both randomly generated graphs and DIMACS \cite{dimacs} graphs. 
For random graphs in our experiments we generated graphs with $|V|=\{50, 100, 150, 200\}$ vertices and
for each we randomly generated edges (uniformly) between vertices so
that the density of edges, $d$, was one of $\{0.1,0.2,0.3, 0.4, 0.5\}$; for each pair of $(|V|, d)$ 10 random instances 
were generated. Therefore, for random graphs,  there is a total of 200 graph instances being generated. 

When assigning color restrictions to 
each vertex we considered lists restricted to five different color ranges.  That is, the randomly
assigned colors were drawn from the range of colors $[1,c|V|]$ where
$c=\{0.1,0.2,0.3, 0.4, 0.5\}$.  Finally, for each color range we considered
three list lengths $k=\{3,4,5\}$. Hence for each graph instance, there are 15 list instances.

For each unique $(|V|,d,c,k)$
tuple there are 10 run instances (each $(|V|, d)$ corresponds to 10 graphs). The four algorithms, $k$-GL, LC, ELC and \texttt{dcc} were run on each instance.  
Both of the two existing heuristic algorithms ran in negligible time. We terminated the branch-and-bound algorithms ELC
and \texttt{dcc} after 1800s (approx.) of elapsed time if it didn't complete the run (in this case, the result is reported as ``$X_n$'').  
For $k$-GL algorithm, in order to benefit from the randomness we run each instance 10 times and then provide the average performance of the run. 
The solution quality of the four algorithms on random graphs\footnote{Though for each pair of $(|V|, d)$ there are 10 graphs, here in the table each 
row corresponds to 1 graph only, since from the experimental data, the performance of each algorithm on each of the 10 graphs are quite similar} with $|V| = \{50, 100 \}$ are reported in Table \ref{tab:random50100} below. The data based on graphs with $|V| = \{150, 200\}$ is eliminated as the performance of the four algorithm over large graphs are quite similar with those in Table \ref{tab:random50100}. 

For DIMACS\cite{dimacs} graphs, we picked all graph instances (approx. 50) from the website, and for each graph instance, there are 15 list instances accordingly.
To facilitate the analysis, we picked the selected set of the instances where $|V| \leq 100$, where details of the 6 graphs is given in Table \ref{tab:dimacs_set}, and the solution quality of the four algorithms on this set
is reported in Table \ref{tab:dimacs} below. 

In Tables \ref{tab:random50100} and \ref{tab:dimacs} blank refers to no solution 
    can be found by the algorithm, `n/s' refers to no feasible solution exists, `$X_n$' refers to a solution of value $X$ found 
   by the algorithm in 1,800s but the search didn't complete .
   \begin{table}
  \small
  \centering
    \caption{Details of the DIMACS \cite{dimacs} graphs where $|V| \leq 100$.}
  \begin{tabular}{|l|l|c|c|c|} \hline
	Tab. \ref{tab:dimacs} name &DIMACS name&$|V|$&$d$&avg deg. / std. dev.\\ 
	\hline
	g1&david.col&87&0.11& 9.33/10.49\\
	\hline
	g2&huck.col&74&0.11& 8.14/7.34\\
	\hline
	g3&jean.col&80&0.08& 6.35/6.02\\
	\hline
	g4&queen8\_12.col&96&0.30& 28.5/2.29\\
	\hline
	g5&queen8\_8.col&64&0.36&22.75/1.85\\
	\hline
	g6&queen9\_9.col&81&0.33&26.07/2.09\\
	\hline
  \end{tabular}
    \label{tab:dimacs_set}
\end{table}
\begin{table}
  \small
  \centering
    \caption{Preliminary comparison on random graphs with $|V| = \{50, 100\}$.}
  \begin{tabular}{|c|c|c|c|c|c|c||c|c|c|c|c||c|c|c|c|c|c|} \hline
  	\multicolumn{17}{|c|}{$|V| = 50$} \\
  	\hline
	d&$|C|$&$\chi _{k=3}$& LC & $k$-GL & ELC& \texttt{dcc} & $\chi_{k = 4}$ & LC & $k$-GL & ELC & \texttt{dcc} & $\chi_{k=5}$&LC&$k$-GL&ELC&\texttt{dcc}\\ 
	\hline
	0.1&0.1&4&&5.0(0.0) & 4& 4 & 4& & 5.0(0.0) & 4 & 4 & 3 &&5.0(0.0)&3&3\\ 
	\hline
	0.1&0.2&6&  & 10.0(0.0) & 6& 6 & 5 &  & 10.0(0.0) & 5 & 5 & 5&&10.0(0.0)&5&5\\ 
	\hline
	0.1&0.3&8&  & 15.0(0.0) & 9& 9 & 7&  & 15.0(0.0) & 7 & 7 & 6&7&15.0(0.0)&6&6\\ 
	\hline
	0.1&0.4&10& & 19.0(0.0) & 12& 10 &8 &  & 18.0(0.0) & 9 & 8 & 7&&20.0(0.0)&8&7\\ 
	\hline
	0.1&0.5&11& & 23.0(0.0) & 13& 12 &10& 10 & 22.0(0.0) & 11 & 10 & 7&&25.0(0.0)&9&7\\ 
	\hline
	0.2&0.1&n/s&  & & & &5&  &  & 5 & 5 & 5&&5.0(0.0)&5&5\\ 
	\hline
	0.2&0.2&8&  & 10.0(0.0) & 8& 8 &7 &  & 10.0(0.0) & 7 & 7 & 6&&10.0(0.0)&6&6\\ 
	\hline
	0.2&0.3&10&  & 15.0(0.0) & 10& 10 &8 &  &15.0(0.0)& 9 & 8 & 7&&10.0(0.0)&7&7\\ 
	\hline
	0.2&0.4&11&  & 19.0(0.0) & 13& 11 & 10 &  & 20.0(0.0) & 11 & 10 & 8&10&20.0(0.0)&8&8\\ 
	\hline
	0.2&0.5&13&  & 24.0(0.0) & 15& 14 & 11& 14 & 21.0(0.0) & 13 & 12 & 9&&22.0(0.0)&10&9\\ 
	\hline
	0.3&0.1&n/s&  & & &  & n/s&  &  &  &  & n/s &&& & \\ 
	\hline
	0.3&0.2&10&  & & 10& 10 & 8 &  & 10.0(0.0)& 8 & 8 & 8&&10.0(0.0)&8&8\\ 
	\hline
	0.3&0.3&12&  & 15.0(0.0) & 12& 12 & 10 &  & 15.0(0.0) & 10 & 10 & 9&&15.0(0.0)&9&9\\ 
	\hline
	0.3&0.4&13&  & 20.0(0.0) & 13& 13 & 11&  & 20.0(0.0) & 11 & 11 & 10&&19.0(0.0)&10&10\\ 
	\hline
	0.3&0.5&14&  & 23.0(0.0) & 16& 14 & 13&  & 23.0(0.0) & 13 & $13_n$ & 11&14&24.0(0.0)&12&$12_n$\\ 
	\hline
	0.4&0.1&n/s&  &  & &  & n/s&  &  &  &  & n/s&&&&\\ 
	\hline
	0.4&0.2&n/s&  &  & &  & 10 &  && 10 & 10 & 9&&&9&9\\ 
	\hline
	0.4&0.3&14&  &  & 14& 14 & 12 &  &  & 12 & 12 & 10&15&15.0(0.0)&10&10\\ 
	\hline
	0.4&0.4&15&  & 19.0(0.0) & 15& 15 & 13 &  & 20.0(0.0)& 13 & 13 & 11&&19.0(0.0)&12&$12_n$\\ 
	\hline
	0.4&0.5&16&  & 23.0(0.0) & 16& 16 & 14 &  & 25.0(0.0) & 15 & $15_n$ &12&&25.0(0.0)&13&$14_n$\\ 
	\hline
	0.5&0.1&n/s&  &  & &  & n/s &  &  &  &  & n/s&&&&\\ 
	\hline
	0.5&0.2&n/s&  &  & &  & n/s &  &  &  &  & n/s&&&&\\ 
	\hline
	0.5&0.3&n/s&  &  & &  & 13 &  &  & 13 & 13 &11&&&11&11\\ 
	\hline
	0.5&0.4&17&  & 20.0(0.0) & 17& 17 & 15 &  &20.0(0.0)& 15 & 15 & 13&16&20.0(0.0)&$13_n$&$14_n$\\ 
	\hline
	0.5&0.5&18&  & 25.0(0.0) & 18& 18 & 15 &  & 25.0(0.0) & $16_n$ & $16_n$ & 14&&25.0(0.0)&$15_n$&$15_n$\\ 
	\hline
	  	\multicolumn{17}{|c|}{$|V| = 100$} \\
  	\hline
	0.1&0.1&9&&10.0(0.0) & 9& 9 & 7& & 10.0(0.0) & 8 & 7 & 7 &&10.0(0.0)&7&7\\ 
	\hline
	0.1&0.2&13&  & 20.0(0.0) & 15& 13 & 11 &  & 20.0(0.0) & 13 & 12 & 9&&20.0(0.0)&11&$10_n$\\ 
	\hline
	0.1&0.3&17&  & 29.0(0.0) & 18& 18 & 14&  & 30.0(0.0) & 17 & 16 & 12&14&30.0(0.0)&14&$14_n$\\ 
	\hline
	0.1&0.4&20&25 & 40.0(0.0) & 27& 27 &15 & 20 & 36.0(0.0) & 21 & 19 & 14&18&37.0(0.0)&20&$16_n$\\ 
	\hline
	0.1&0.5&23& & 43.0(0.0) & 37& 33 &18& 22 & 44.0(0.0) & 27 & $23_n$ & 15&&43.0(0.0)&23&$20_n$\\ 
	\hline
	0.2&0.1&n/s&  & & & &10&  &  & 10 & 10 & 9&&&10&$10_n$\\ 
	\hline
	0.2&0.2&16&  & 20.0(0.0) & 17& 16 &14 &  & 20.0(0.0) & 15 & $14_n$ & 12&&20.0(0.0)&13&$13_n$\\ 
	\hline
	0.2&0.3&20& 26 & 30.0(0.0) & 23& $21_n$ &17 &  &30.0(0.0)& 21 & $20_n$ & 14&19&30.0(0.0)&17&$17_n$\\ 
	\hline
	0.2&0.4&23&  & 39.0(0.0) & 31& $26_n$ & 19 &  & 40.0(0.0) & 25 & $23_n$ & 17&20&39.0(0.0)&22&$21_n$\\ 
	\hline
	0.2&0.5&26&  & 47.0(0.0) & 32& $31_n$ & 21&  & 47.0(0.0) & 29 & $27_n$ & 18&&48.0(0.0)&27&$25_n$\\ 
	\hline
	0.3&0.1&n/s&  & & &  & n/s&  &  &  &  & n/s &&& & \\ 
	\hline
	0.3&0.2&n/s&  & & &  & 17 &  && 18 &$17_n$ & 15&&20.0(0.0)&$16_n$&$16_n$\\ 
	\hline
	0.3&0.3&24&  & & 25& $24_n$ & 19 &  & 30.0(0.0) & $22_n$ & $22_n$ & 17&&30.0(0.0)&$20_n$&$20_n$\\ 
	\hline
	0.3&0.4&27&  & 39.0(0.0) & 30& $31_n$ & 23&  & 40.0(0.0) & $28_n$ & $27_n$ & 19&24&39.0(0.0)&$24_n$&$24_n$\\ 
	\hline
	0.3&0.5&30&  & 46.0(0.0) & 34& $33_n$ & 24 &  & 46.0(0.0) & $29_n$ & $29_n$ & 21&25&49.0(0.0)&$29_n$&$29_n$\\ 
	\hline
	0.4&0.1&n/s&  &  & &  & n/s&  &  &  &  & n/s&&&&\\ 
	\hline
	0.4&0.2&n/s&  &  & &  & 20 &  && 20 & tout & 18&&&$19_n$&$19_n$\\ 
	\hline
	0.4&0.3&28&  &  & 28& 28 & 23 &  &30.0(0.0)  & $25_n$ & $27_n$ & 20&&30.0(0.0)&$23_n$&$23_n$\\ 
	\hline
	0.4&0.4&31&  & 39.0(0.0) & 33& $33_n$ & 25 &  & 40.0(0.0)& $30_n$ & $31_n$ & 23&29&40.0(0.0)&$29_n$&$29_n$\\ 
	\hline
	0.4&0.5&34&  & 48.0(0.0) & $38_n$& $36_n$ & 27 &  & 48.0(0.0) & $34_n$ & $35_n$ &24&32&49.0(0.0)&$31_n$&$33_n$\\ 
	\hline
	0.5&0.1&n/s&  &  & &  & n/s &  &  &  &  & n/s&&&&\\ 
	\hline
	0.5&0.2&n/s&  &  & &  & n/s&  &  &  &  & 26&&&&\\ 
	\hline
	0.5&0.3&n/s&  &  & &  & 27 &  &  & $28_n$ &$28_n$ &23&&30.0(0.0)&$27_n$&$27_n$\\ 
	\hline
	0.5&0.4&35&  & & 35& $36_n$ & 30 &  &40.0(0.0)& $33_n$ & $34_n$ & 26&&40.0(0.0)&$33_n$&$33_n$\\ 
	\hline
	0.5&0.5&37&  & 49.0(0.0) & $39_n$& $40_n$ & 30 &  & 48.0(0.0) & $37_n$ & $38_n$ & 28&&49.0(0.0)&$36_n$&$37_n$\\ 
	\hline
  \end{tabular}
  \label{tab:random50100}
\end{table}

The implementation of all algorithms were written in \texttt{C++} and
were complied using \texttt{gcc} version 4.7.2.  All experiments were conducted
on Dell desktops with the software/hardware configuration as Fedora release 18 (Spherical Cow), 3.3GHz Intel Core i5-2500
processor with 8GB of DDR3 memory (clock speed 1333MHz).

\begin{table}[H]
  \small
  \centering
    \caption{Preliminary comparison on DIMACS graphs with $|V| \leq 100 $. }
  \begin{tabular}{|c|c|c|c|c|c|c||c|c|c|c|c||c|c|c|c|c|c|} \hline
	name &$|C|$&$\chi _{k=3}$& LC & $k$-GL & ELC& \texttt{dcc} & $\chi_{k = 4}$ & LC & $k$-GL & ELC & \texttt{dcc} & $\chi_{k=5}$&LC&$k$-GL&ELC&\texttt{dcc}\\ 
	\hline
	g1&0.1&n/s&  & & & &n/s&  &  &  &  & n/s&&&&\\ 
	\hline
	&0.2&13&  & 17.0(0.0) & 13& 13 &12 &  &17.0(0.0) & 12 & 12 & 11&&17.0(0.0)&11&11\\ 
	\hline
	&0.3&16& & 26.0(0.0) & 16& 16 &14 &  &26.0(0.0)& 14 & $15_n$ & 13&&26.0(0.0)&$14_n$&$14_n$\\ 
	\hline
	&0.4&17&  & 31.0(0.0) & 18& 17 & 15 &  & 32.0(0.0) & $16_n$ & $17_n$ & 13&&32.0(0.0)&$14_n$&$15_n$\\ 
	\hline
	&0.5&20&  & 40.0(0.0) & 22& $24_n$ & 17&20  & 40.5(1.6) & $20_n$ & $20_n$ & 16&&38.0(0.0)&$18_n$&$18_n$\\ 
	\hline
	g2&0.1&n/s&  & & & &n/s&  &  &  &  & n/s&&&&\\ 
	\hline
	&0.2&13&  & & 13& 13 &11 &  &14.0(0.0) & 11 & 11 & 11&&14.0(0.0)&11&11\\ 
	\hline
	&0.3&14& & 22.0(0.0) & 14& 14 &12 &  &22.0(0.0)& 12 & $12_n$ & 11&&22.0(0.0)&11&$12_n$\\ 
	\hline
	&0.4&15&  & 27.0(0.0) & 16& $16_n$ & 13 &  & 27.0(0.0) & $14_n$ & $13_n$ & 12&&27.0(0.0)&$12_n$&$14_n$\\ 
	\hline
	&0.5&17&  & 33.0(0.0) & $18_n$& $17_n$ & 14&  & 35.0(0.0) & $15_n$ & $15_n$ & 13&&31.0(0.0)&$14_n$&$15_n$\\ 
	\hline
	g3&0.1&n/s&  & & & &n/s&  &  &  &  & n/s&&&&\\ 
	\hline
	&0.2&12&  & 16.0(0.0)& 12& 12 &10 & 13 &16.0(0.0) & 10 & 10 & 10&&16.0(0.0)&10&10\\ 
	\hline
	&0.3&13& & 24.0(0.0) & 13& 13 &12 &  &24.0(0.0)& 12 & $12_n$ & 11&&24.0(0.0)&11&$12_n$\\ 
	\hline
	&0.4&15&  & 28.0(1.0) & 15& $17_n$ & 12 &  & 32.0(0.0) &12 & $14_n$ & 12&17&30.0(0.0)&$13_n$&$13_n$\\ 
	\hline
	&0.5&18& 21 & 34.0(0.0) & 19& $19_n$ & 14&  & 37.0(0.0) & $17_n$ & $17_n$ & 13&&39.0(0.0)&$15_n$&$16_n$\\ 
	\hline
	g4&0.1&n/s&  & & & &n/s&  &  &  &  & n/s&&&&\\ 
	\hline
	&0.2&n/s&  & & & &16 & &19.0(0.0) & $17_n$ & $17_n$ & 15&&19.0(0.0, 1\footnotemark)&$16_n$&$16_n$\\ 
	\hline
	&0.3&21& & 28.0(0.0) & $22_n$& $23_n$ &19 &  &28.0(0.0)& $22_n$ & $23_n$ & 17&&28.0(0.0)&$20_n$&$20_n$\\ 
	\hline
	&0.4&26&  & 36.0(0.0) & $29_n$& $28_n$ & 22 &  & 37.0(0.0) &$26_n$ & $25_n$ & 19&&37.0(0.0)&$24_n$&$26_n$\\ 
	\hline
	&0.5&28& 34 & 46.0(0.0) & $32_n$& $31_n$ & 23&  & 45.0(0.0) & $30_n$ & $30_n$ & 21&&46.0(0.0)&$27_n$&$28_n$\\ 
	\hline
	g5&0.1&n/s&  & & & &n/s&  &  &  &  & n/s&&&&\\ 
	\hline
	&0.2&n/s&  & & & &12 & & & 12 & 12 & 11&&&11&$11_n$\\ 
	\hline
	&0.3&18& &  & 18& 18&14 &  &19.0(0.0)& 14 & $15_n$ & 12&&19.0(0.0)&$14_n$&$14_n$\\ 
	\hline
	&0.4&19&  & 24.0(0.0) & 19& $19_n$ & 15 &  & 25.0(0.0) &17 & $17_n$ & 14&19&25.0(0.0)&$15_n$&$16_n$\\ 
	\hline
	&0.5&21& & 32.0(0.0) & 23& $23_n$ & 16&  & 30.0(0.0) & $20_n$ & $19_n$ & 15&21&31.0(0.0)&$18_n$&$18_n$\\ 
	\hline
	g6&0.1&n/s&  & & & &n/s&  &  &  &  & n/s&&&&\\ 
	\hline
	&0.2&16&  & &16 &16 &15 & & & $15_n$ & $15_n$ & 13&&&$14_n$&$15_n$\\ 
	\hline
	&0.3&19& & 24.0(0.0) & 20& $20_n$&16 &  &24.0(0.0)& $18_n$ & $19_n$ & 15&&24.0(0.0)&$17_n$&$17_n$\\ 
	\hline
	&0.4&22&  & 31.0(0.0) & 25& $24_n$ & 18 &  & 32.0(0.0) &$21_n$ & $22_n$ & 17&&31.0(0.0)&$20_n$&$20_n$\\ 
	\hline
	&0.5&25& & 40.0(0.0) & 28& $28_n$ & 20&  27& 40.0(0.0) & $24_n$ & $26_n$ & 18&&39.0(0.0)&$24_n$&$23_n$\\ 
	\hline
  \end{tabular}
  \footnotetext{Footnote}

    \label{tab:dimacs}
\end{table}

\subsection{Evaluation of the $k$-GL and LC}\label{sec:heur-evaluation} 

Tables \ref{tab:random50100} and \ref{tab:dimacs} show the solution quality of the two 
existing heuristic algorithms, $k$-GL and LC, as well as the two newly developed ones, ELC and \texttt{dcc}, on
both random graphs with $|V| = \{50, 100 \}$ and DIMACS graphs with $|V| \leq 100$.

In view of the performance of each
algorithm, not surprisingly, \textit{k}-GL provides a poorer solution
quality than \textit{LC} if the latter finds a feasible solution.  But  it's surprising 
to see that $k$-GL succeed in finding a feasible
solution over the 85\% of the instances in a single run as shown in the tables. 
  More surprising, though, is the poor showing
 of LC, the independent set-based heuristic.  Of the 190 instances where feasible solutions
 exist the algorithm failed to find a solution (shown as empty entries
 below where ``$\chi_{k=x}$'' is not ``n/s'') 163 times (over 85\%). These failures, generally, correlate with small
 color ranges. 
 
We conducted another experiment additionally to investigate the poor performance of LC phenomenon.  
We considered graphs with
$|V|=\{50,100\}$ vertices and $d$ one
of $\{$0.1, 0.2, 0.3, 0.4, 0.5$\}$.  For the color availability list the randomly 
assigned colors were drawn from the range of colors $[1,c|V|]$ where $c=\{1.0, 0.9, 0.8\}$
and the list length $k = \{3, 4, 5\}$. Each pair of $(|V|, d, k)$ is a run, therefore for each $c$ 
there are 30 runs. 
In this experiment LC managed to find a feasible 
solution over 29 of the 30 runs when $c = 1.0$, 27 when $c=0.9$, and 25 when $c=0.8$
This result, along with the figure of 85\% failure in Tables \ref{tab:random50100} and \ref{tab:dimacs} 
 suggest that for smaller color ranges, there is a higher chance that the independent set based heuristic will
fail in finding a feasible solution. 

In summary, the results shown in Tables \ref{tab:random50100} and \ref{tab:dimacs} confirm the trends below
\begin{itemize}
\item for greater value of list length, $k$, a smaller number of
  colors are required to list color a graph, generally;
\item for smaller graph densties, a smaller number of colors is needed
  to color the graph;
\item in general fewer colors are needed when the color range becomes
  smaller.
\end{itemize}

\subsection{\textit{ELC} and \texttt{dcc} Evaluation} \label{sec:elc}

Tables \ref{tab:random50100} and \ref{tab:dimacs} compare the solution quality of the two
newly developed branch-and-bound based algorithm,  ELC and \texttt{dcc} on both random graphs with $|V| = \{50, 100 \}$ and DIMACS graphs
with $|V| \leq 100$.

Again, the three tables show the experimental results of each algorithm on a total of 250 run instances. Among 
 the 250 instances there are 60 instances where feasible solution doesn't exist (shown as ``n/s'' under ``$\chi_{k=x}$'' ).
The two algorithms, ELC and \texttt{dcc}, managed to find a feasible solution in 189 of the 190 instances where 
feasible solutions exist. 

In regard to the solution quality, when $|V| = 50$ ELC managed to reach the optimal result over 39 of the 60 
instances where feasible solutions exist, and \texttt{dcc} managed over 49 of the 60; when $|V| = 100$, ELC found 5 while \texttt{dcc} found 
10 of 60. For the set of DIMACS graphs, ELC found 32 optimal solutions while \texttt{dcc} found 26 of 70. For g1, g2, g3 in 
Table \ref{tab:dimacs_set}, the edges in the graphs are distributed unevenly and this may somehow explains why the performance
of the algorithms on g1, g2, and g3 is not consistent with the performance on random graphs where $|V| = 100$ and $d = 0.1$, though they have a similar $(|V|, d)$. 

The best solution found by \texttt{dcc} is always better or equal to the one found by ELC when both completed the run.
When $|V| = 50$, both the two algorithms found the same result  on 45 of the 60 instances. For the remaining 15 instances, \texttt{dcc} won 13 while ELC won 2(\texttt{dcc}
didn't complete the run on the two graphs). When $|V| = 100$, there are 25 instances with a tie, for the remaining 35 instances, \texttt{dcc} won 25 while ELC won 10(again, for the 10 graphs, \texttt{dcc} didn't complete the run). Meanwhile, in comparison with LC, \texttt{dcc} provides better results than LC for nearly all the graphs that \texttt{dcc} completed the run and LC found a feasible solution.  

In summary, the results shown in Tables \ref{tab:random50100} and \ref{tab:dimacs} confirm the trends below
\begin{itemize}
\item for greater value of list length -- $k$, more running time is
required to list color a graph, generally; 
\item for greater graph densties, more running time is needed
  to color the graph.
\end{itemize}

Further, in view of the performance of ELC and \texttt{dcc},  when $|V| = 100$ ELC fail to complete the run on over 38\% of the 60 instances where feasible solutions exist, and dcc fails on 77\% of the 60 instances, not even considering the cases of when $|V| = \{150, 200\}$.  To facilitate the improvement of exact
algorithms, it is reasonable, we suggest, that further research focus on
$|V|$ between $[50, 100]$ as a threshold to narrow down the area.
 
\section{Conclusions}\label{sec:concls} %

We have implemented and investigated the only two existing list coloring
algorithms known to us in the literature and have proposed a
branch-and-bound based algorithm with additional data structure support
through the use of a priority queue. The performance of each has been
investigated and reported on.

Further algorithm tuning opportunities exist: there will be at least one initial feasible clique coloring
that could lead to the optimal solution eventually, and it may need more than 5,000 iterations
to reach the final optimal coloring during the BB search, therefore it would be nice if such 
initial clique coloring can be identified and then the BB search based on the `good' initial coloring is 
given more time to compute a tighter bound. This, and also identifying other opportunities for 
code optimization are our next priorities.

\section*{Acknowledgments}\label{sec:ack} %
The research is jointly funded by the \textit{Irish Research Council}
(IRC), and the \textit{China Scholarship Council} (CSC). 
Thanks are also due to three anonymous referees for feedbacks in improving 
the paper, CSIS Department for providing the experimental facilities, Fen Chen and 
Yijun Yin for assistance in setting up the experiments.

\bibliography{top}
\bibliographystyle{plain}
\end{document}